\documentclass[article,accept,moreauthors,pdftex,10pt,a4paper,preprints]{mdpi} 

\firstpage{1} 

\history{}

\usepackage[utf8]{inputenc}
\usepackage[T1]{fontenc}
\usepackage{hyperref}
\usepackage{graphicx,bm,amsmath,color,scalerel}
\usepackage{bbold}
\usepackage{wasysym}
\usepackage{verbatim}
\usepackage{physics}
\usepackage{subcaption}
\graphicspath{ {./img/} }
\usepackage{nomencl}
\makenomenclature
\usepackage{etoolbox}
\renewcommand\nomgroup[1]{%
  \item[\bfseries
  \ifstrequal{#1}{Z}{Abbreviations}{%
  \ifstrequal{#1}{N}{Number Sets}{%
  \ifstrequal{#1}{O}{Other Symbols}{}}}%
]}

\newcommand{\be}{\begin{equation}}
\newcommand{\ee}{\end{equation}}
\newcommand{\beq}{\begin{eqnarray}}
\newcommand{\eeq}{\end{eqnarray}}
\newcommand{\ba}{\begin{align}}
\newcommand{\ea}{\end{align}}

\Title{Lorentz violation footprints in the spectrum of high-energy cosmic neutrinos: \\
Deformation of the spectrum of superluminal neutrinos from electron-positron pair production in vacuum}

\Author{José Manuel Carmona$^{1}$*\orcidA{}, José Luis Cortés$^{1}$\orcidB{}, José Javier Relancio$^{1}$\orcidC{} and Maykoll A. Reyes$^{1}$\orcidD{}}

\AuthorNames{José Manuel Carmona, José Luis Cortés, José Javier Relancio and Maykoll A. Reyes}

\address{%
$^{1}$ \quad Departamento de F\'{\i}sica Te\'orica and Centro de Astropartículas y Física de Altas Energías (CAPA),
Universidad de Zaragoza, Zaragoza 50009, Spain; jcarmona@unizar.es (J.M.C); cortes@unizar.es (J.L.C.); relancio@unizar.es (J.J.R.); maykollreyes.phys@gmail.com (M.A.R.)}

\corres{Correspondence: jcarmona@unizar.es}

\abstract{The observation of cosmic neutrinos up to 2 PeV is used to put bounds on the energy scale of Lorentz invariance violation through the loss of energy due to the production of $e^+e^-$ pairs in the propagation of superluminal neutrinos. A model to study this effect, which allows to understand qualitatively the results of numerical simulations, is presented.}


\begin{document}

\section{Introduction}
\label{sec:introduction}

Special relativity (SR) postulates Lorentz invariance as an exact symmetry of Nature. It is at the base of our quantum field theories of the fundamental interactions, and has surpassed all experimental tests up to date (\cite{Kostelecky:2008ts,Long:2014swa,Kostelecky:2016pyx,Kostelecky:2016kkn}; see also the papers in~\cite{LectNotes702}). In general relativity (GR), local Lorentz invariance still holds, but not as a global symmetry of spacetime. When considering a curved spacetime, the global symmetries are given by the isometries of the metric characterizing the curvature~\cite{Weinberg:1972kfs}. However, it is not clear how to introduce symmetries of spacetime in a quantum gravity theory (QGT), since spacetime plays a completely different role in GR and in quantum field theory (QFT). In QFT, a particular spacetime is given and one studies the properties and interactions of particles on it, but in GR spacetime appears as a dynamical variable affected by the material content. 

One possibility is that Lorentz invariance is indeed broken for high enough energies. This is studied in Lorentz invariance violation (LIV) scenarios (see Ref.~\cite{Liberati2013} for a review), usually formulated within the theoretical framework known as the standard model extension (SME), an effective field theory in which new terms that violate Lorentz invariance are added to the usual terms appearing in the Lagrangian of the standard model (SM)~\cite{Colladay:1998fq}. The existence of a violation of Lorentz invariance implies a privileged system of reference, for which the cosmic background radiation is usually assumed to be isotropic.

In this way, Lorentz symmetry would be only a good long-distance, or low-energy approximation to the true symmetries of a QGT, that should be parametrized by a certain high-energy scale $\Lambda$. This scale is supposed to be of the order of the Planck mass, $m_P\approx 1.2\times 10^{19}$\,GeV/$c^2$, whose direct exploration is certainly out of reach in present experiments. However, over the past few years it has been realized that there are astrophysical observations that could be sensitive to such deviations~\cite{Mattingly:2005re}. For example, thresholds of reactions can be significantly changed by modifications of SR in such a way that processes that are kinematically forbidden in SR could become allowed for a certain energy, much lower than the Planck mass. Also, the possible existence of an energy dependent velocity for massless particles could be observed in measurements of the time of arrival of photons emitted by far way sources, 
since there is an amplification effect due to the long distance they travel.

Another possible observable effect could appear in the  spectrum of the detected neutrinos in the IceCube experiment~\cite{Aartsen:2013jdh,Aartsen:2014gkd,Aartsen:2018fqi}. In particular, there seems to be a cutoff in the spectrum for energies of the order of some PeV, which is in contrast with an extrapolation of the energy dependence of the flux for energies up to 60 TeV. Such an extrapolation, together with the presence of the Glashow resonance at 6.3 PeV~\cite{Glashow:1960zz} (for which an electronic anti-neutrino interacts with an electron at rest producing a real $W^{-}$ boson), would predict the detection of a few events at these energies that have not been observed. Although more data have to be collected in order to assure the existence of a cutoff, one can speculate about this suppression. Then, two options can be considered: either there is some kind of mechanism at the source in such a way that high energy neutrinos are not emitted with the usual law for lower energies, or there is an effect of new physics that has to be taken into account.

This is the proposal of Ref.~\cite{Stecker:2014oxa}. In this work, a LIV scenario was considered, so that new processes forbidden in SR are now kinematically allowed. In particular, they considered the processes of neutrino splitting (NS) and vacuum electron-positron pair emission (VPE). With a Monte Carlo analysis, they found that dimensions 6 operators produce a cutoff in the spectrum of detected neutrinos.

In this work, we find the same cutoff in the spectrum of neutrinos but, instead of considering a Monte Carlo simulation, we will use an analytic method that follows the propagation of neutrinos by considering their energy loss due to the universe expansion and the VPE effect. The NS process is not considered in this analytic method, since it involves a non-conservation of the number of neutrinos. The inclusion of this effect will require to follow the evolution of the full neutrino spectrum rather than of individual neutrinos and is left for a future work.

In Section \ref{sec:flux}, we will model the flux of detected and emitted neutrinos for one source, establishing a relation between them. This relation between fluxes will depend on the relation between the emitted energy and the detected energy for each neutrino, which is computed in Section \ref{sec:energyloss}. Finally, we will merge both results in Section \ref{sec:spectrum}, in order to obtain a prediction for the detected flux, knowing the characteristics of the emission and the distribution of sources. We conclude in Section \ref{sec:conclusions} with a discussion of the very stringent bounds on the scale $\Lambda$ of LIV that one gets from the observation by IceCube of cosmic neutrinos up to 2 PeV.

\section{Neutrino flux} 
\label{sec:flux}

In order to relate the emitted neutrinos at $z_e$ with the detected neutrinos at $z=0$, we will consider a Friedman-Lemaître-Robertson-Walker (FLRW) model for the expanding Universe, for which the redshift is defined from the evolution of the scale factor,
\be
a(t)=\frac{a_0}{1+z},
\ee
where $a_0$ is the scale factor at $z=0$. From the previous expression, and introducing the Hubble parameter $H(t)=\dot{a}(t)/a(t)$, one gets the relation between $dt$ and $dz$
\be
dt\,=\,-\frac{dz}{H(z)(1+z)} \,,
\label{eq:dtdz}
\ee
while the FLRW equation gives
\begin{equation}
  H^2(z)=H_0^2\qty[\Omega_m\,(1+z)^3+\Omega_r\,(1+z)^4+\Omega_\Lambda+\Omega_\kappa\,(1+z)^2] \,,
  \label{eq:hz}
\end{equation}
with $H_0$ the value of the Hubble parameter today, and $\Omega_m$, $\Omega_r$, $\Omega_\Lambda$, $\Omega_\kappa$, the density fractions of matter, radiation, dark energy and curvature respectively, with values~\cite{PhysRevD.98.030001}
\begin{equation}\begin{gathered}
  \Omega_\Lambda \sim 0.692 \pm 0.012\qquad
  \Omega_m \sim 0.308 \pm 0.012\\
  \Omega_\kappa \sim 0.005 \pm 0.017\qquad
  \Omega_r \sim 5.38 \cdot 10^{-5} \pm 0.0015 \,.
\end{gathered}\end{equation} 
Neglecting the contributions of the density fractions of the radiation and curvature, one gets  
\begin{equation}
H(z)\approx H_0 \sqrt{\Omega_m\,(1+z)^3+\Omega_\Lambda} \;.
  \label{eq:H(z)}
\end{equation}
The comoving distance between the emission and detection points of a neutrino is related to $z_e$ by
\be
r_e(z_e)=-\int \frac{dt}{a(t)}=\int_0^{z_e} \frac{dz}{a_0 H(z)},
\ee
where we have used the relation $ds^2=0=dt^2-a^2(t)dr^2$ from the FLRW metric and Eq.~\eqref{eq:dtdz}. Then the emitted neutrinos at $z_e$ are spread in an area $4\pi a_0^2 r_e^2(z_e)$ at $z=0$. 

In order to determine the flux of the detected neutrinos, we need to start from a model for the origin of the high energy neutrinos. This model can be summarized in a function $P_e(E_e,z_e)$, which gives the number of emitted neutrinos $dN_e$ in a time interval $\delta t_e$, with energies in the interval $(E_e, E_e+dE_e)$, and at a distance from the detector corresponding to redshifts in the interval $(z_e,z_e+dz_e)$, as
\be
    dN_e(E_e,z_e) \,=\, P_e(E_e,z_e) \, dE_e dz_e \delta t_e\,.
    \label{eq:emitted_nu}
\ee

Neutrinos emitted at redshift $z_e$ with an energy $E_e$ arriving to the detector will be detected with a lower energy $E_d$, which results from the energy loss in the propagation of the neutrinos. Since the processes of energy loss that we are going to consider in this work do not change the number of neutrinos during their propagation from its source to the detector, there exists a one-to-one correspondence between a neutrino detected with energy $E_d$ and the emission at a certain $z_e$ of a neutrino with energy $E_e$, which will be a function of $E_d$ and $z_e$. Let us call $g(E_d,z_e)$ to this function, that is, $E_e=g(E_d,z_e)$. The determination of the function $g$ will be the objective of Sec.~\ref{sec:energyloss}. It allows us to write Eq.~\eqref{eq:emitted_nu} in terms of the detected energy:
\be
dN_e(g(E_d,z_e),z_e) \,=\, P_e(g(E_d,z_e),z_e) \pdv{g(E_d,z_e)}{E_d} \, dE_d dz_e \frac{\delta t_d}{1+z_e}\,,
\label{eq:emitted_nu-2}
\ee
where we have used that an interval $\delta t_e$ at emission is stretched at detection by a factor $(1+z_e)$, which is the ratio of scale factors at the source and at the detector.
 
From Eq.~\eqref{eq:emitted_nu-2}, one can get the number of neutrinos with energies in the interval $(E_d, E_d+dE_d)$, which arrive in a time interval $\delta t_d$ to an area $\delta A$ subtended by the detector from the source, and then, the spectral neutrino flux (measured in convenient units as number of neutrinos per $\text{GeV}\cdot \text{s}\cdot \text{cm}^2 \cdot \text{str}$) at an energy $E_d$ will be
\be
\phi_d(E_d)\equiv \frac{dN_d(E_d)}{dE_d \delta t_d \delta A}=\int dz_e P_e(g(E_d,z_e),z_e) \pdv{g(E_d,z_e)}{E_d} \frac{1}{1+z_e} \frac{1}{4\pi a_0^2 r_e(z_e)^2}\,.
\ee

The function $P_e(E_e,z_e)$ depends on the model for the emission of the neutrinos. If we take the simple model that the neutrinos are emitted from sources which follow a density distribution $\rho(z_e)$, according to a power law $E_e^{-\alpha}$, $\alpha=2$,\footnote{This is the approximation made in Ref.~\cite{Stecker:2014oxa} from the data of IceCube; however, one could go beyond that approximation by including the neutrinos generated in the interstellar medium by cosmic ray interactions (diffuse flux)~\cite{Carceller:2017tvc}.} then
\be
    P_e(E_e,z_e) \,\propto\, \frac{1}{E_e^2} \rho(z_e) \,,
		\label{modelo}
\ee
where one can consider the scenario in which the redshift distribution of the sources $\rho(z_e)$ approaches that of the star formation rate~\cite{Stecker:2014xja}. If $C$ is the constant proportionality factor in Eq.~\eqref{modelo}, the detected spectral neutrino flux for this model of emission of neutrinos will be
\be
\phi_d(E_d) \,=\, C \int dz_e \,\rho(z_e) \frac{1}{g^2(E_d,z_e)} \pdv{g(E_d,z_e)}{E_d} \frac{1}{1+z_e} \frac{1}{4\pi a_0^2 r_e(z_e)^2}\,.
\ee

\section{Energy loss in the propagation of a superluminal neutrino}
\label{sec:energyloss}

In this section, we will consider the differential evolution of the neutrino energy along the trajectory, which will give us the relation $g(E_d,z_e)$ between the emitted and detected energies of a neutrino propagating from the source to the detector. This will be determined by the classical effect of the expansion of the universe, on the one hand, and by the effect of the VPE, the effect of new physics, on the other hand. In order to get the total energy variation due to both effects, we will analyze each of them independently.

\subsection{Expansion of the Universe}

The variation of the neutrino energy due to the expansion of the Universe is well known from the dilation of the wavelength or the contraction of the frequency
\be 
   \nu_d\,=\,\frac{\nu}{(1+z)} \,.
\ee
Recalling that for an (approximately) massless particle, $E= h\nu$, we get the relation 
\be
    E_d\,=\,\frac{E}{(1+z)} \,.
    \label{eq:ed_ee}
\ee
This gives, for a fixed detected energy, the neutrino energy as a function of $z$. Now, differentiating the previous relation, we obtain $dE=E_d\,dz$. Substituting here Eq.~\eqref{eq:ed_ee}, we get the differential variation of the neutrino energy due to the expansion of the Universe:
\be
    \frac{dE}{E}\,=\,\frac{dz}{(1+z)} \,.
    \label{eq:var_expansion}
\ee

\subsection{Vacuum Pair Emission}

The second mechanism of energy loss is the emission of electron-positron pairs through the process \mbox{$\nu_i\rightarrow \nu_i + e^- + e^+$}. In the process of VPE, a superluminal neutrino with an energy-momentum relation 
\be
E^2 \,=\, p^2  + \frac{p^{2+n}}{\Lambda^n}
\ee 
can produce two new particles when its energy is above a threshold energy $E^*$. This threshold energy is given in terms of the energy scale $\Lambda$ and the order of the correction $n$ by~\cite{Carmona2012}:
\be
    E^*\,=\,\qty(4m_e^2\Lambda^n)^{1/(2+n)} \,.
    \label{eq:e_star}
\ee

\begin{figure}[h!tb]
    \centering
    \begin{subfigure}{0.5\textwidth}
        \centering
        \includegraphics[width=\textwidth]{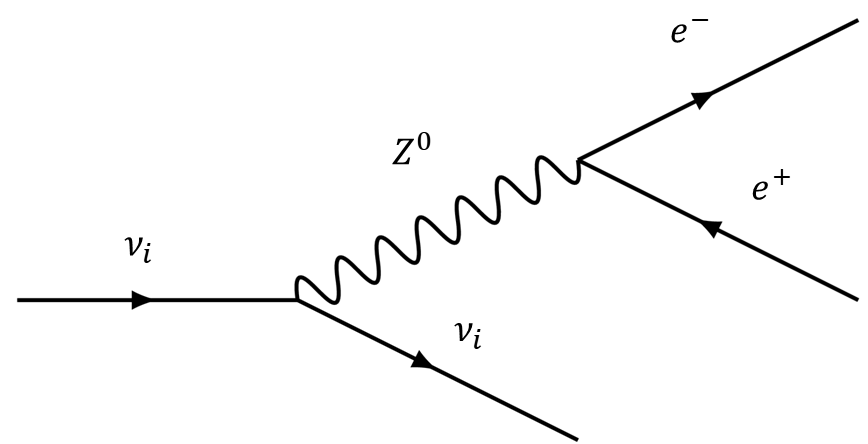}
        \caption{Neutral channel}
        \label{fig:vpe_z}
    \end{subfigure}%
    \begin{subfigure}{0.5\textwidth}
        \centering
        \includegraphics[width=\textwidth]{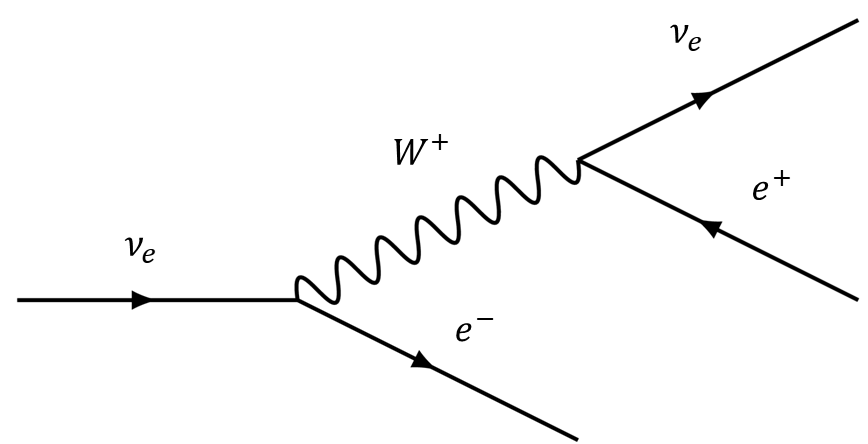}
        \caption{Charged channel}
        \label{fig:vpe_w}
    \end{subfigure}
    \caption{Neutrino disintegration through VPE}
\end{figure}

The disintegration can be produced through two different ways: a neutral channel, mediated by a boson $Z^0$, and a charged channel, mediated by a boson $W^+$ (pictures \ref{fig:vpe_z} and \ref{fig:vpe_w}, respectively).
However, one should consider the charged channel as an additional contribution to the neutral channel only for electron neutrinos, that is, roughly 1/3 of the time during superluminal propagation, due to the phenomenon of neutrino oscillations. As a first approximation, we can neglect it and consider the neutral channel as the dominant one. With this simplification, the process has been characterized in Ref.~\cite{Carmona2012}:
\be
    \Gamma \,=\, \frac{G_F^2p^5}{192\pi^3}\qty[(1-2s_W^2)^2+(2s_W)^2]\qty(\frac{p}{\Lambda})^{3n} \xi_n \,,
\ee
where $G_F$ is the Fermi constant, $s_W=\sin(\theta_W)$ is the sine of the Weinberg angle, and $\xi_n$ is a constant (dependent of $n$) of order 1. From this decay width, in the same reference, it is also obtained the variation of the momentum in time
\be
    \dv{p}{t} \,=\, \frac{G_F^2p^6}{192\pi^3}\qty[(1-2s_W^2)^2+(2s_W^2)]\qty(\frac{p}{\Lambda})^{3n} \xi_n' \,.
\ee

Noting that for massless neutrinos, $dp/dt=dE/dt$, we can write
\be
    \frac{dE}{dt}\,=\,-\alpha_n E^{6+3n} \,,
\ee
where we have defined $\alpha_n$ as
\be
    \alpha_n \,=\, \frac{G_F^2\xi_n'}{192\pi^3\Lambda^{3n}}\qty[(1-2s_W^2)^2+(2s_W)^2] \;.
    \label{eq:alpha_n}
\ee

Using the relation between $dt$ and $dz$ given by Eq.~\eqref{eq:dtdz}, 
we get the evolution of the neutrino energy due to the VPE
\be
    \frac{dE}{E}\,=\,\frac{\alpha_n E^{5+3n}dz}{H(z)(1+z)} \,.
\ee
 
\subsection{Evolution in case of VPE}

As the process of VPE has a threshold energy, when the neutrino energy goes down below that energy, pair production stops. Let us assume  that the VPE has been occurring between points $z_i$ and $z_f$ of the trajectory. In that case, the kinematics of the propagation of the neutrinos between those points is determined by 
\begin{equation}
    \frac{dE}{E}\,=\, \frac{dz}{(1+z)} + \frac{\alpha_n E^{5+3n}\,dz}{H(z)(1+z)} \,,
    \label{eq:dos_efectos}
\end{equation}
where the first term on the right hand side takes into account the expansion and the second one, the VPE. In order to determine $E$ as a function of $z$, we start by defining  $\widetilde E\equiv E/(1+z)$: 
\begin{equation}
  E\,=\,\widetilde E (1+z) \implies dE\,=\,d\widetilde E \;(1+z)+ \widetilde E\,dz \,.
\end{equation}
Substituting in Eq.~\eqref{eq:dos_efectos}, one finds
\begin{equation}
  \frac{d\widetilde E \;(1+z)+ \widetilde E\;dz}{\widetilde E (1+z)}\,=\,
  \frac{dz}{(1+z)} + \frac{\alpha_n \widetilde E^{5+3n}(1+z)^{5+3n}\,dz}{H(z)(1+z)}\,,
\end{equation}
and then 
\begin{equation}
    \frac{d\widetilde E}{\widetilde E}\,=\,\frac{\alpha_n \widetilde E^{5+3n}(1+z)^{4+3n}\,dz}{H(z)} \,.
\end{equation}
Using Eq.~\eqref{eq:H(z)} and defining $y\equiv (1+z)^3$, then $dy=3(1+z)^2dz$, and therefore
\begin{equation}
  \frac{d\widetilde E}{\widetilde E^{6+3n}}\,=\,\frac{\alpha_n}{3H_0}\;\frac{y^{2/3+n}}{\sqrt{\Omega_m\,y+\Omega_\Lambda}}\,dy \,.
  \label{eq:vpe_tilde}
\end{equation}

Now, integrating from the initial point $z_i$, where the neutrino has an energy $E_i$, to the final point $z_f$, where the neutrino has an energy $E_f$, one gets
\begin{equation}
  \frac{1}{(5+3n)}\qty(\widetilde {E_f}^{-(5+3n)}-\widetilde E_i^{-(5+3n)}) \,=\,
  \frac{\alpha_n}{3H_0}\,\int_{(1+z_f)^3}^{(1+z_i)^3}\frac{y^{2/3+n}}{\sqrt{\Omega_m\,y+\Omega_\Lambda}}\,dy \,.
  \label{eq:integral_general}\end{equation} 
	From this expression, one can write the initial energy as 
	\begin{equation}
  E_i=\frac{(1+z_i)}{(1+z_f)}E_f\qty(1+(5+3n)\frac{\alpha_n}{3H_0}J_n(z_i,z_f)\left(\frac{E_f}{1+z_f}\right)^{5+3n})^{-\frac{1}{(5+3n)}} \,,
  \label{eq:E_i}
\end{equation}
where we have defined 
\begin{equation}
J_n(z_i,z_f)\equiv \int_{(1+z_f)^3}^{(1+z_i)^3}\frac{y^{2/3+n}}{\sqrt{\Omega_m\,y+\Omega_\Lambda}}\;dy \,.
  \label{eq:j(z)}
\end{equation}

Now, substituting Eq.~\eqref{eq:alpha_n} (definition of $\alpha_n$) in Eq.~\eqref{eq:E_i}, and using Eq.~\eqref{eq:e_star} (definition of $E^*$) to express $\Lambda$ as a function of $E^*$, we can rearrange the expression of the emitted energy in the following way
\begin{equation}
  E_i=\frac{(1+z_i)}{(1+z_f)}E_f\qty(1-\lambda_n(z_i,z_f) \left(\frac{E_c}{E^*}\right) \left(\frac{E_f}{E^*}\right)^{5+3n})^{-\frac{1}{(5+3n)}} \,,
  \label{Ei-Ef}
\end{equation}
where we have factorized the dependence on $z_i,z_f$ introducing the quantity
\be
\lambda_n(z_i,z_f)\, \equiv \,\frac{(5+3n)  \xi'_n}{3}\qty[(1-2s_W^2)^2+\qty(2s_W)^2]\frac{J_n(z_i,z_f) }{(1+z_f)^{5+3n}} \,,
\ee
and the dependence on energy by defining an energy scale $E_c$
\be 
    E_c\equiv\frac{G_F^2 \left(4 m_e^2\right)^3}{192\,\pi^3\,H_0} \,, 
\ee
which is of order EeV.

\section{Deformation of the spectrum of high-energy neutrinos}
\label{sec:spectrum}

Having identified in the previous section the neutrino energy loss due to the production of $e^+e^-$ pairs in the propagation on an expanding universe, we study in this section the bounds on the energy scale ($\Lambda$) of LIV that one can get from the observations of cosmic neutrinos by IceCube. 

If we have a neutrino detected with an energy $E_d$ greater than the threshold $E^*$ for $e^+e^-$ pair production in the decay of a neutrino, then one can apply the relation (\ref{Ei-Ef}) 
with $z_i=z_e$, $E_i=E_e$, $z_f=0$ and $E_f=E_d$. Then, the energy $E_e$ of a neutrino emitted at a point with a redshift $z_e$ arriving to the detector at $z=0$ with an energy $E_d$ is
\be
E_e \,=\, (1+z_e)\,E_d\,\left[1-\lambda_n(z_e,0)\,(E_c/E^*)\,(E_d/E^*)^{5+3n}\right]^{-1/(5+3n)} \,\equiv\, g^{(1)}(E_d,z_e)\,.
\ee
 From this expression one can see that introducing $z_e^{*(1)}(E_d)$ by the condition 
\be
\lambda_n(z_e^{*(1)}(E_d), 0) \,=\, \left(\frac{E^*}{E_c}\right) \,\left(\frac{E^*}{E_d}\right)^{5+3n}\,,
\ee
one has that the neutrino has to be emitted with $z_e < z_e^{*(1)}(E_d)$, otherwise the neutrino can not arrive to the detector with an energy $E_d > E^*$. Taking into account that $E_c\sim \text{EeV}$, that all observations of cosmic neutrinos are at energies much smaller than $E_c$, and that $(E^*/E_d)^{5+3n}\ll 1$ unless $E_d$ is extremely close to $E^*$, one has $z_e^{*(1)}(E_d)\ll 1$.

Only neutrinos emitted from points very close to the detector ($z_e\ll 1$) can arrive to the detector with an energy $E_d>E^*$ and then, independently of the details of the model for the origin of the cosmic neutrinos, one will have a strong suppression of the spectrum for $E_d>E^*$. Explicitly, one has 
\be
 \phi_d(E_d)  \,=\, C \,\int_0^{z_e^{*(1)}(E_d)} \,dz_e \,\rho(z_e) \frac{1}{g^{(1)}(E_d,z_e)^2} \pdv{g^{(1)}(E_d,z_e)}{E_d} \frac{1}{1+z_e} \frac{1}{4\pi a_0^2 r_e(z_e)^2}\,.
\label{eq:first_integral}
\ee 

Next, one can consider a neutrino detected with an energy $E_d$ smaller than $E^*$. We introduce 
\be
z^*(E_d) \equiv \frac{E^*}{E_d} - 1\,,
\ee
so that neutrinos emitted from $z_e < z^*(E_d)$ cannot produce $e^+e^-$ pairs. Then, in this case, the change in the energy of the neutrino in the propagation is just due to the expansion of the universe $E_e = E_d (1+z_e)$. If the neutrino is emitted from $z_e > z^*(E_d)$, then one can apply the relation (\ref{Ei-Ef}) with $z_i=z_e$, $E_i=E_e$, $z_f=z^*(E_d)$ and $E_f=E^*$
\be
E_e \,=\, (1+z_e)\,E_d\,\left[1-\lambda_n(z_e,z^*(E_d)) \,(E_c/E^*)\right]^{-1/(5+3n)} \,\equiv\, g^{(2)}(E_d,z_e)\,. 
\ee
Once more, one can introduce a critical value $z_e^{*(2)}(E_d)$ by the condition 
\be
\lambda_n(z_e^{*(2)}(E_d), z^*(E_d)) = \left(\frac{E^*}{E_c}\right)\,,
\ee
so that one has that the neutrino is emitted from $z^*(E_d) < z_e < z_e^{*(2)}(E_d)$. The neutrinos with $E_d<E^*$ which are affected by a suppression due to LIV effects in the propagation are those which are emitted from $z_e > z^*(E_d)$. Then 
\begin{align}
\phi_d(E_d)  &\,=\, C\,\int_{z^*(E_d)}^{z_e^{*(2)}(E_d)} \,dz_e \,\rho(z_e) \frac{1}{g^{(2)}(E_d,z_e)^2} \pdv{g^{(2)}(E_d,z_e)}{E_d} \frac{1}{1+z_e} \frac{1}{4\pi a_0^2 r_e(z_e)^2} \nonumber \\&+\, C\int_{0}^{z^*(E_d)} \,dz_e\,\rho(z_e) \frac{1}{E_d^2 (1+z_e)^2}\frac{1}{4\pi a_0^2 r_e(z_e)^2}\,,
\label{eq:second_integral}  
\end{align}
and the suppression in the spectrum will decrease when $E_d$ decreases. 

The strong suppression of the spectrum of detected neutrinos at energies $E_d > E^*$ implies that from the observations of cosmic neutrinos by IceCube extending beyond a few PeV, we can conclude that the threshold energy ($E^*$) for $e^+e^-$ production  should be of the order of or greater than a few PeV. The exact bound on $E^*$ (and then on $\Lambda$) will depend on the details of the model for the origin of the cosmic neutrinos. 

As in Ref.~\cite{Stecker:2014oxa}, we will assume, as an illustrative example,  that the neutrino sources have a redshift distribution similar to that of the star formation rate~\cite{Behroozi:2012iw}. In particular, we will consider sources from $z=0.2$ to $z=6.8$. Let us start with the first integral of Eq.~\eqref{eq:first_integral}, that takes into account the neutrinos detected with an energy higher  than the threshold energy, which we will consider to be $10\,\text{PeV}$ in order to have a suppression in the neutrino spectrum due to LIV in the range of energies accessible to IceCube. It is not difficult to show that one would not detect such neutrinos, since the value of  $z_e^{*(1)}(E_d)$ is lower than the redshift of the nearest considered source,  $z=0.2$. To illustrate that, one can see that for the particular case of $E_d=E^*$, one finds $z_e^{*(1)}(E^*)=4\times 10^{-5}$, and this value will get smaller for higher energies, so there is a total suppression to detect neutrinos with $E_d>E^*$. 

\begin{figure}[p]
    \centering
        \includegraphics[width=0.9\textwidth]{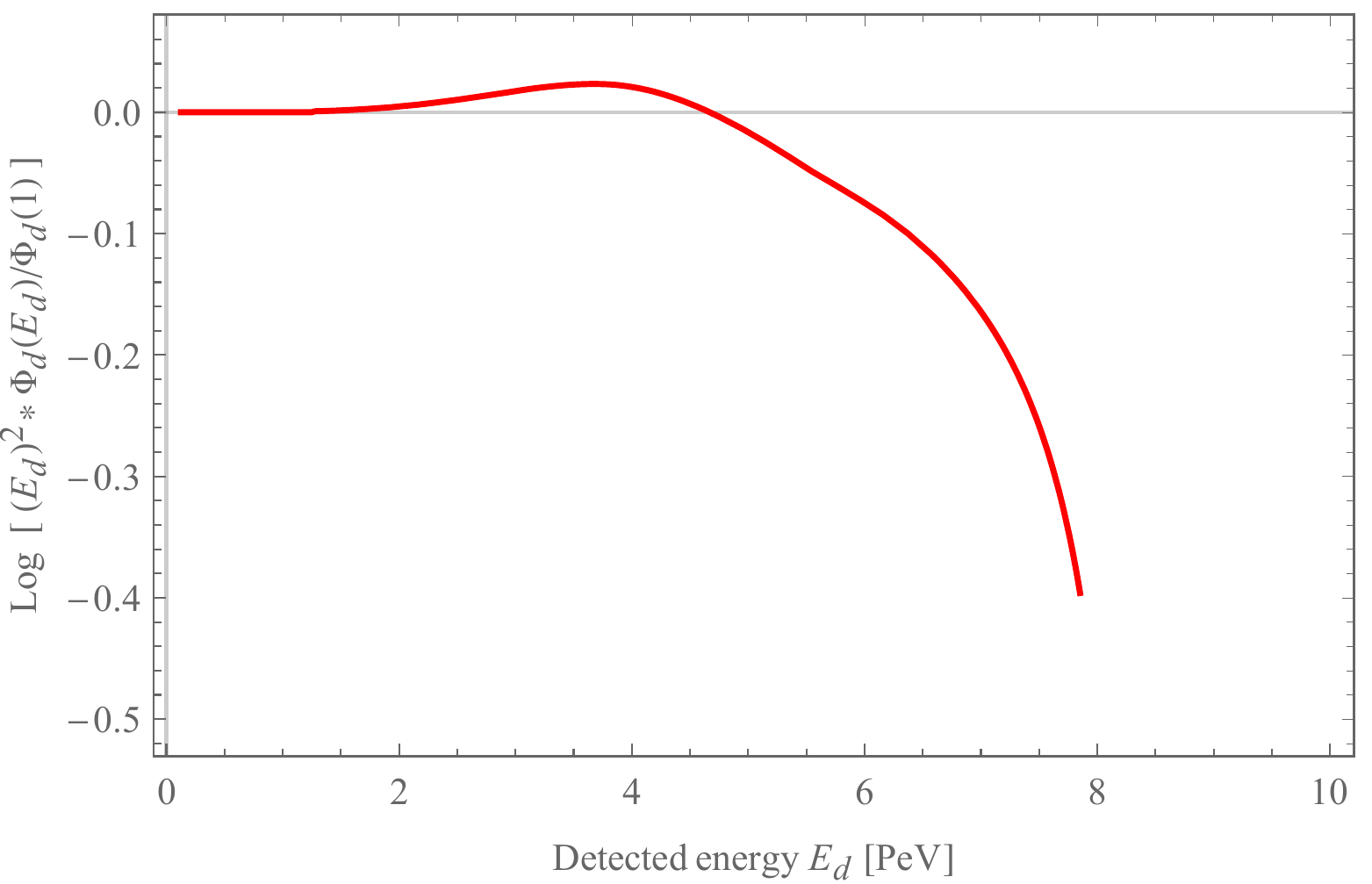}
				 \caption{Logarithmic representation of the flux of neutrinos for $n=1$.}
				\label{fig:1}
\end{figure}

\begin{figure}[p]
    \centering
        \includegraphics[width=0.9\textwidth]{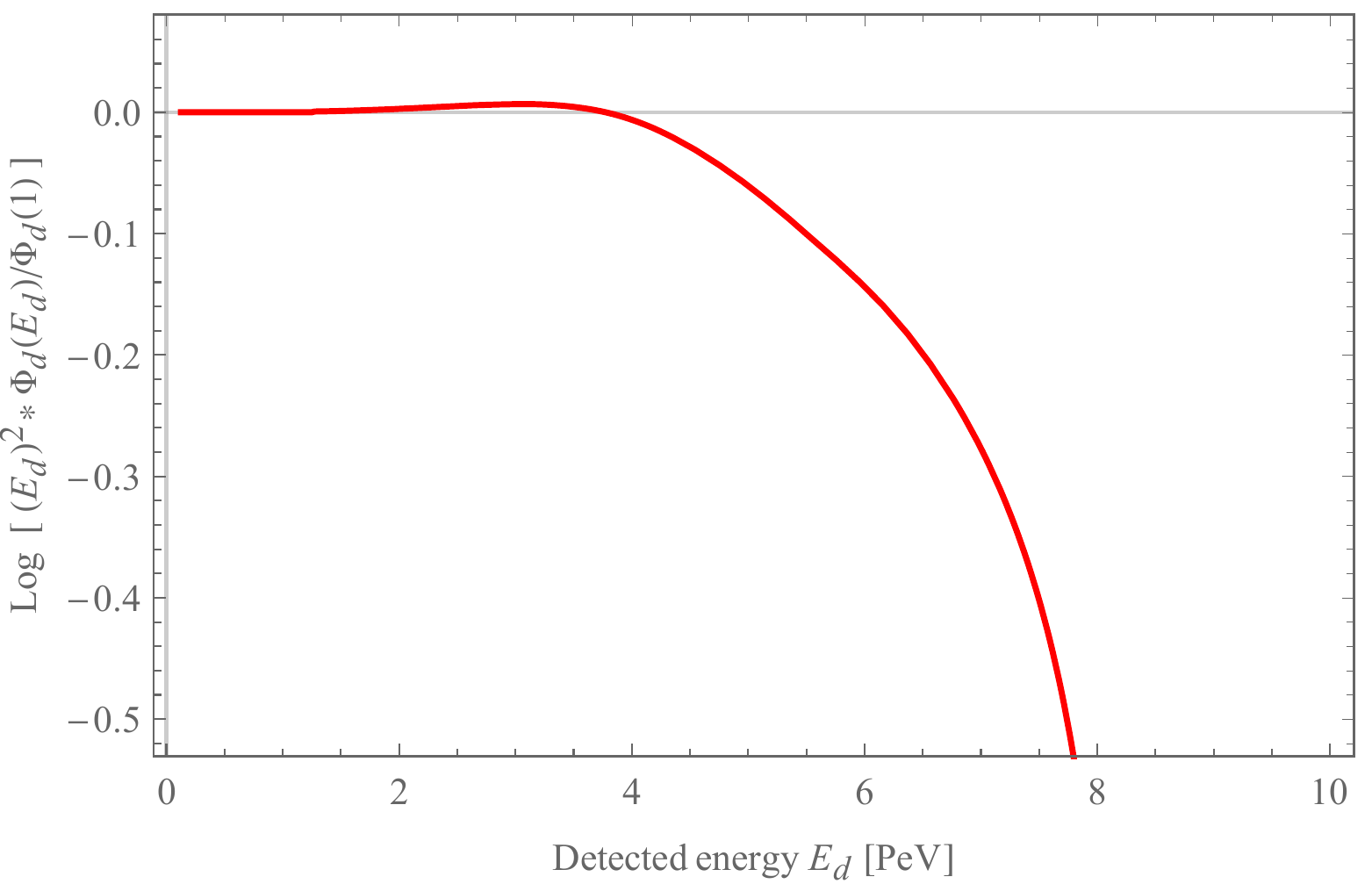}
				 \caption{Logarithmic representation of the flux of neutrinos for $n=2$.}
					\label{fig:2}
\end{figure}

For the second integral, Eq.~\eqref{eq:second_integral}, we obtain the normalized distribution for $n=1$ and for $n=2$ represented in Figs.~\ref{fig:1} and \ref{fig:2}, respectively. 

We find in both cases a cutoff in the spectrum of neutrinos below the threshold energy, preceded by a small bump. This is in fact the same result obtained in Ref.~\cite{Stecker:2014oxa} with a Monte Carlo simulation when only VPE was considered. 

From the cutoff, one could extract a constrain to $\Lambda$ for this model in order to be compatible with the neutrino observations until $2\, \text{PeV}$.  If one forgets the neutrino splitting effect, this model could give an explanation of the absence of detected neutrinos for energies above   $2\,\text{PeV}$ despite the Glashow resonance at  $6.3\,\text{PeV}$, imposing that the rapid fall of the probability of detection occurs between  $2\,\text{PeV}$ and the resonance energy $6.3\,\text{PeV}$.

Since the last detected neutrino of the spectrum has an energy of $2\, \text{PeV}$, we know from our model that the  threshold energy has to be greater than this value. From this, we can get the minimum value of $\Lambda$ for $n=1$ and $n=2$ from Eq.~\eqref{eq:e_star}, obtaining  $7.6\times 10^{18}\, \text{PeV}$ and $3.9\times 10^{9}\, \text{PeV}$ respectively. While in the first case the high energy scale is five orders of magnitude grater than the Planck energy, in the second one is four orders below. These bounds on $\Lambda$ should be compared with the less stringent bounds obtained from other possible observable effects of LIV, like time delays in Gamma Ray Bursts (GRBs) or the end of the ultra-high-energy cosmic-ray (UHECR) spectrum. 

The conclusions obtained with this simple model for the origin of the neutrinos can be easily extended to a more realistic model that takes into account a more trustworthy punctual source distribution, together with a diffuse component due to the production of neutrinos in interactions of ultra energetic cosmic rays in the gas of our galaxy, in other galaxies, or in the intergalactic medium~\cite{Carceller:2017tvc}. All these contributions will change the detected spectrum in a small range of energies below the scale $E^*$, where the  rapid fall of the spectrum occurs.

\section{Discussion and conclusions}
\label{sec:conclusions}

We have used the detection of high energy neutrinos as an example where a breaking of Lorentz invariance with an energy scale much larger than the energy scale of any observation can lead to observable effects, thanks to the amplification due to the propagation of a particle over very large distances.  The best candidate for the particle is a neutrino, due to the very weak interaction which allows to consider a free propagation. If one has a modification of the energy-momentum relation of SR such that the energy of a particle with a given momentum is increased due to the breaking of Lorentz invariance, then, for sufficiently high energies, the lightest neutrino becomes unstable since it can decay through weak interactions into a neutrino and an $e^+e^-$ pair (one could also have a $\mu^+\mu^-$ pair or a $\tau^+\tau^-$ pair, but the higher masses of the charged leptons make the threshold energy of the decay far beyond any accessible energy, except in the case of $e^+e^-$). We have considered the relation between the energy of a neutrino at the source and at the detector due to the effect of the expansion of the universe, and the loss of energy in the production of $e^+e^-$ in the propagation of the neutrino between the source and the detector. 

Given this relation between the emitted and detected energies, one can determine the distortion of the neutrino spectrum due to the breaking of LIV for a given model for the spectrum of emission of neutrinos and the distribution of sources. Such distortion produces a very step fall in the flux of neutrinos in a range of energies below the threshold of the neutrino decay. We have determined the range of energies where this fall in the neutrino flux occurs using a simplified model for the origin of the high energy neutrinos.  The observation of cosmic neutrinos up to $2\,\text{PeV}$ implies that the threshold of the decay has to be larger than a few PeV and this translates in a very stringent lower bound on the energy scale $\Lambda$ of LIV. 

There is another possible decay of neutrinos due to weak interactions with the production of a neutrino-antineutrino pair (neutrino splitting). In this case, one has a cascade of neutrinos associated with each emitted neutrino instead of a single neutrino and one has to go beyond the model presented in this work. The distortion of the neutrino spectrum can not be obtained in this case from a model for the evolution of the energy of a neutrino in the propagation from the source to the detector. A model allowing to incorporate the effects of all the decays of neutrinos will be the subject of a future work. In any case, this is an effect on top of the effect due to the production of $e^+e^-$ pair, which will not invalidate the bound obtained in this work but will replace it by still a more stringent constrain on the scale $\Lambda$ of LIV.  

\authorcontributions{All authors contributed equally to the present work.}

\acknowledgments{This work is supported by the Spanish grants FPA2015-65745-P (MINECO/FEDER), PGC2018-095328-B-I00 (FEDER/Agencia estatal de investigación), and by the Spanish DGIID-DGA Grant No. 2015-E24/2. The authors would also like to thank support from the COST Action CA18108.}

\conflictsofinterest{The authors declare no conflict of interest.} 







\externalbibliography{yes}
\bibliography{QuGraPhenoBib}

\end{document}